\documentclass[12pt,a4paper]{article}
\usepackage[margin=1in]{geometry}
\usepackage[utf8]{inputenc}
\usepackage[english]{babel}
\usepackage{amsmath,amsfonts,amssymb,graphicx,amsthm,authblk}
\usepackage{physics}
\usepackage[T1]{fontenc}
\usepackage[utf8]{inputenc}

\usepackage[english]{babel}

\usepackage[colorinlistoftodos]{todonotes}
\usepackage[colorlinks=true, allcolors=blue]{hyperref}

\title{Moyal Star-Product and Unitary Representations of the Euclidean Motion Group}

\author{Alexander J. Balsomo and Job A. Nable}

\date{}

\begin{document}
\maketitle\large

\begin{abstract}
In this paper, the Moyal star-product quantization is used to construct the unitary irreducible representations of the Euclidean motion group on 3-dimensions. These unitary representations will come from the representation of its Lie algebra whose operators are defined by the left Moyal star-product multiplication. In fact, these representations of the Lie algebra is the infinitisimal representation. Hence, the exponentiation of these operators gives rise to unitary operators that defines the desired unitary representations. 
\end{abstract}

\section{Introduction}
The mathematical formulation of unitary representations of the Euclidean motion group $\mathrm{M}(3)$ is widely known and the earliest accounts of its exposition are found in \cite{Miller,Vilenkin}.  As a semidirect product of the compact group $\mathrm{SO}(3)$ with the abelian group $\mathbb{R}^3$, the construction of this group's unitary irreducible representation fits within the framework of induced representation \cite[Theorem 7.7]{Berndt} and this is illustrated in \cite{Chirikjian}.

On the other hand, the quantum theory of a free particle provides a recipe to construct the unitary representations of symmetry groups.  This is well-known in the relativistic case of the Poincar\'{e} group but less known in the non-relativistic case of $\mathrm{M}(3)$ \cite{Woit}. As a Lie subgroup of the Jacobi group, the restriction of the Schr\"{o}dinger representation to $\mathrm{M}(3)$ paves the way for the construction of the unitary representation of the said group.  It is known that the derived Schr\"{o}dinger representation  leads to the quantization of only at most degree 2 polynomials \cite{Groenewold}.

As an autonomous quantization theory, deformation quantization \cite{Bayen} suggests the introduction of a noncommutative but associative product, called the star-product ($\star$-product), on the space of $C^\infty$-function on a symplectic manifold to construct a model of quantum mechanics; hence, quantum mechanics is a deformed algebra of classical observables which retains the key ingredient of quantization-the correspondence principle.  As in the case of Kirillov's orbit method \cite{Kirillov}, the papers   \cite{Arnal1,Arnal2,Arnal3} constructed and classified the unitary irreducible representations of the nilpotent and exponential Lie groups by methods of $\star$-product quantization, essentially utilizing the Moyal $\star$-product.  

Star-product quantization is an important aspect of the Weyl-Wigner-Groenewold-Moyal formalism or the phase space formalism of quantum mechanics. The quasiprobability distributions given by the Wigner functions may be obtained as matrix elements of the unitary irreducible representations of the Heisenberg group. It may be viewed as the Weyl transform \cite{Wong} of the projection operator $\rho= \ket\psi\bra\psi$ corresponding to a pure state, and generalized to arbitrary operators. This generalization has an inverse called the Weyl quantization, still the most popular of the several quantization methods available. The Weyl transform of the composition of two Weyl operators results precisely in the Moyal star-product \cite{Moyal} of phase space functions. 

The method presented in the papers by V\'{a}rilly, Gracia-Bond\'{i}a and their coworkers is the reverse to the general procedure presented in this work \cite{Varilly1,Varilly2}. These works exhibit the construction of a noncommutative product on the space of functions on the phase space, induced from the operator product of unitary operators coming from the projective representations of the invariance group of the quantum system. The idea has been developed so far as to implement harmonic analysis on phase space and to derive special function identities \cite{Varilly3,Varilly4}.

The authors of this paper have outlined the procedure in \cite{Balsomo} with the Euclidean motion group $\mathrm{M}(2)$ as an example.  This motion group of 2 dimension is solvable and we have shown that the Moyal $\star$-product on the space of classical observables on a cylinder- a coadjoint orbit of $\mathrm{M}(2)$, generates the operators that define the unitary representation of this group. A crucial point is that the construction is dependent on coordinatization of the cylinder. This is to be expected, as quantization is highly coordinate dependent \cite{Poirier}. Following this outline, we will construct the unitary representations of the nonsolvable group $\mathrm{M}(3)$. We make the important remark that the computations here are essentially elementary and explicit which can be read by students of mathematics and physics with little background in Lie Theory and Quantum Mechanics. Moreover, the recipe discussed in the next section seem to work for large classes of Lie groups, certainly for the general case of $\mathrm{M}(n),$ which will be considered elsewhere.

We have organized the paper as follows.  Section 2 presents the unitary representation of the Eucliden motion group $\mathrm{M}(3)$.  In Section 3, we outline the $\star$-product method of constructing unitary representations. The construction of the unitary representations of $\mathrm{M}(3)$ via the Moyal $\star$-product is shown in Section 4 and we present our conclusion in the last section.

\section{Unitary Representations of $\mathrm{M}(3)$}

The Euclidean motion group $\mathrm{M}(3)$ is identified with the multiplicative group of matrices of size 4 of the form
\begin{equation}
\left(\begin{array}{cc}
R & r \\
0 & 1
\end{array}\right)
\end{equation}
where $R$ is a rotation in $\mathrm{SO}(3)$ and $r$ is a vector in $\mathbb{R}^3$.  This group's corresponding Lie algebra $\mathfrak{m}(3)$ is the space of matrices of the form 
\begin{equation}
\left(\begin{array}{cc}
A & v \\
0 & 0
\end{array}\right)
\end{equation}
where $A$ is a skew-symmetric matrix in $\mathfrak{so}(3)$ and $v$ a vector in $\mathbb{R}^3$.  This algebra is spanned by the basis elements $X_i$ for $i=1,2,3$ where
\begin{equation}
X_1=\left(\begin{array}{ccc}
0 & -1 & 0 \\
1 & 0 & 0 \\
0 & 0 & 0
\end{array}\right),\quad
X_2=\left(\begin{array}{ccc}
0 & 0 & 1 \\
0 & 0 & 0 \\
-1 & 0 & 0
\end{array}\right),\quad
X_3=\left(\begin{array}{ccc}
0 & 0 & 0 \\
0 & 0 & -1 \\
0 & 1 & 0
\end{array}\right)
\end{equation}	   
of $\mathfrak{so}(3)$ and the canonical basis $E_1,E_2$ and $E_3$ of $\mathbb{R}^3$.  The group elements of $\mathrm{M}(3)$ relates with the vectors of $\mathfrak{m}(3)$ by
\begin{equation}
(R,r)=\exp(r_1E_1+r_2E_2+r_3E_3)\exp(\theta_1X_1)\exp(\theta_2X_2)\exp(\theta_3X_3)
\end{equation}
where $r=(r_1,r_2,r_3)\in\mathbb R^3$ and $\theta_1,\theta_2,\theta_3$ the Euler angles representing the matrix $R$.  These $\exp(r_iE_i)$ and $\exp(\theta_jX_j)$ are the 1-parameter subgroups of $\mathrm{M}(3)$ where $i,j=1,2,3$. 

We denote the matrices above as $g=(R,r)$ and $U=(A,v)$. The class 1 unitary representation $\mathcal{U}:\mathrm{M}(3)\to\mathrm{Unit}(L^2(S^2))$ of $\mathrm{M}(3)$ is defined by the operators \cite{Vilenkin}
\begin{equation}\label{UnitRep}
(\mathcal{U}^\lambda_gf)(s)=\mathrm{e}^{\mathrm{i}\lambda r\cdot s}f(R^{-1}s),
\end{equation}
where $\lambda>0$, $f$ a square-integrable function on the $2$-sphere and $\cdot$ the standard inner product over $\mathbb{R}^3$. The set $\{\mathcal{U}^\lambda:\lambda>0\}$ is the complete collection of infinite-dimensional unitary irreducible representations of $\mathrm{M}(3)$. The representation $\mathrm{d}\mathcal{U}^\lambda$, defined by the derivative
\begin{equation}\label{LinearRep}
\mathrm{d}\mathcal{U}^\lambda(U)=\left.\frac{\mathrm{d}}{\mathrm{d}t}\mathcal{U}^\lambda_{\exp(tU)}\right|_{t=0}
\end{equation}
is the infinitisimal representation of the Lie algebra $\mathfrak{m}(3)$, associated to the representation $\mathcal{U}^\lambda$ of $\mathrm{M}(3)$.  But this linear representation (\ref{LinearRep}) of the  algebra is completely determined by the operators
\begin{equation}\label{Position}
\mathrm{d}\mathcal{U}^\lambda(E_i)=\mathrm{i}\lambda Q,
\end{equation}
where the operator $Q=$ multiplication of $s_i, s=(s_1,s_2,s_3)\in S^2$ and
\begin{equation}\label{Momentum}
\mathrm{d}\mathcal{U}^\lambda(X_j)=\mathrm{i}P
\end{equation}	
where the operator $P=\mathrm{i}\displaystyle\frac{\partial}{\partial s_j}$.  Hence, the unitary representation of $\mathrm{M}(3)$, defined in (\ref{UnitRep}), is recovered from (\ref{Position}) and (\ref{Momentum}) and is given by the unitary operators 
\begin{eqnarray}\label{BCH Unitary}
\mathcal{U}^\lambda_g&=&\exp(r_1\mathrm{d}\mathcal{U}^\lambda(E_1)+r_2\mathrm{d}\mathcal{U}^\lambda(E_2)+r_3\mathrm{d}\mathcal{U}^\lambda(E_3))\nonumber \\
&&\times\exp(\theta_1\mathrm{d}\mathcal{U}^\lambda(X_1))\exp(\theta_1\mathrm{d}\mathcal{U}^\lambda(X_1))\exp(\theta_1\mathrm{d}\mathcal{U}^\lambda(X_1)).
\end{eqnarray}	 	 
Here, the Taylor series expansion of the exponential of the momentum operator $\mathrm{i}P$ provides the action of translations on the phase space while the exponential of the coordinate operator $\mathrm{i}\lambda Q$ gives the action of rotation. 

In the method of induced representation, the unitary representation of a Lie group is induced from the representation of its closed subgroup and the representation space that makes this representation irreducible are the square-integrable functions on the homogeneous space of cosets associated to this closed subgroup.  For $\mathrm{M}(3)$, its unitary representation (\ref{UnitRep}) is induced from the representation of the subgroup $\mathrm{SO}(2)\ltimes\mathbb{R}^3$ and the square-integrable functions are defined on the homogeneous space $\mathrm{M}(3)/\mathrm{SO}(2)\ltimes\mathbb{R}^3$ which is identified with $S^2$.  On the other hand, another way to construct the unitary representation of $\mathrm{M}(3)$ is via the unitary representation of the Jacobi group $G^J(3)$, and then restricting it to its subgroup $\mathrm{M}(3)$.  Since $\mathfrak{m}(3)$ is a Lie subalgebra of $\mathfrak{g}^J(3)$, quantization via the Schr\"{o}dinger representation provides the Lie algebra representation of $\mathfrak{m}(3)$ and its exponentiation is the desired unitary representation of $\mathrm{M}(3)$ \cite{Woit}.  In fact, the representation in (\ref{BCH Unitary}) is the Fourier-transformed version of the unitary representation computed via the Schr\"{o}dinger representation.  This unitary representation is irreducible on the space of wavefunctions as solution to the time-independent Schr\"{o}dinger equation, with the Casimir operator as its differential operator.    

\section{Outline of the Program}

In this section, we outline the construction of the unitary representation of $\mathrm{M}(3)$. The techniques below were tested for nilpotent Lie groups \cite{Arnal1, Arnal2} and exponential Lie groups \cite{Arnal3}, but some concrete computations were also made with Lie groups that are neither nilpotent nor exponential, to wit: affine transformation of the real and complex plane \cite{Diep}, rotation groups \cite{Nable} and $\mathrm{MD}_4$-groups \cite{Nguyen}. The same technique works well in the construction of unitary irreducible representations of Euclidean motions $\mathrm{M}(2)$ in two-dimensional plane \cite{Balsomo}.

\subsection{Identifying the coadjoint orbits of a Lie group}  	

A coadjoint orbit $\Omega_F$ of a Lie group $G$ is a homogeneous symplectic manifold. It can be expressed as the quotient space $G/G_F$ where $G_F$ is a stabilizer subgroup of $G$ with respect to the coadjoint action of $G$ on a fixed linear functional $F$ in the dual space $\mathfrak{g}^*$ of the Lie algebra $\mathfrak{g}$.  It is an immersed submanifold of $\mathfrak{g}^*$ and carries a symplectic structure  $\omega_F$ called the Kirillov symplectic form, given by the inner product $\langle F,[U,T]\rangle$, for any tangent vector $U,T$ in $\mathfrak{g}$.

For the Euclidean motion group $\mathrm{M}(3)$, the computation has been carried out in \cite[\S 19]{Guillemin} and this is the set
\begin{equation}\label{CoOrbit}
\Omega_F=\{(R\mu+R\alpha\times r,R\alpha): (R,r)\in\mathrm{M}(3)\}
\end{equation}  
where $F=(\mu,\alpha)$ in $\mathfrak{m}(3)^*$ and this dual space of $\mathfrak{m}(3)$ is identified with $\mathbb{R}^3\times\mathbb{R}^3$.  Besides the trivial orbit when $\mu=\alpha=0$, there are families of 2-dimensional and 4-dimensional orbits:  2-spheres of radius $\|\alpha\|$ and the cotangent bundles of the 2-spheres.  

As a consequence of Darboux's theorem, a symplectic manifold is locally flat and hence, for every point $m$ on the coadjoint orbit $\Omega_F$, there symplectomorphically corresponds to a neighborhood $O$ of $m$ a flat subspace of an even-dimensional Euclidean space with the standard symplectic form.  So for an appropriate choice of symplectic coordinates $(p,q)$, this 2-form can be written as $\mathrm{d}q\wedge\mathrm{d}p$ on $O$.  Though the ideal chart on the nontrivial orbit $\Omega_F$ is given by the polar coordinate system, however, because of the nonvanishing higher order bidifferential operators of the Moyal $\star$-product, this global chart will not produce a covariant $\star$-product which will be explained next.  

\subsection{The Hamiltonian system and the covariant Moyal $\star$-product} 

The symplectic manifold $(\Omega,\omega)$ is the natural arena of Hamiltonian mechanics, where the space of $C^\infty$-functions on $\Omega$ models classical mechanics with the ordinary pointwise product and the Poisson bracket $\{\cdot,\cdot\}$ providing its associative and Lie algebra structures, respectively. The vector field $\xi_H$ associated to an energy function $H$ that satisfy the equation $i(\xi_H)\omega=\mathrm{d}H$ is called the Hamiltonian vector field, and the triple $(\Omega,\omega,\xi_H)$ is called the Hamiltonian system. If $(p,q)$ are the canonical coordinates of $\omega$, then the Hamiltonian vector field \cite{Abraham} is expressed as
\begin{equation}
\xi_H=\sum_i\left(\frac{\partial H}{\partial p_i}\frac{\partial}{\partial q_i}-\frac{\partial H}{\partial q_i}\frac{\partial}{\partial p_i}\right)
\end{equation}	 
since the derivative of the energy function is given by
\begin{equation}
\mathrm{d}H=\sum_i\left(\frac{\partial H}{\partial p_i}\mathrm{d}p_i+\frac{\partial H}{\partial q_i}\mathrm{d}q_i \right).
\end{equation}

When the manifold is the coadjoint orbit $\Omega_F$, we define these energy functions $\widetilde{U}:\Omega_F\to\mathbb{R}$ on $\Omega_F$ by the dual pairing
\begin{equation}\label{EnergyFunc}
\widetilde{U}(F')=\langle F',U  \rangle
\end{equation}	   	
of the vector $U$ in $\mathfrak{g}$ and the linear functionals $F'$ in the orbit. So, the collection $\widetilde{\mathfrak{g}}=\{\widetilde{U}:U\in\mathfrak{g}\}$ is a finite-dimensional Lie subalgebra of $(C^\infty(\Omega_F),\{\cdot,\cdot\})$ and for every $U,T$ in $\mathfrak{g}$, this satisfies
\begin{equation}\label{Symplectic}
\omega_F(\xi_U,\xi_T)=\widetilde{[U,T]}
\end{equation}	     
on the fixed $F$.

On $C^\infty(\Omega,\omega)$, we introduce the Moyal $\star$-product of two smooth functions $f$ and $g$, defined by
\begin{equation}
f\star g=fg+\sum_{r=1}^\infty\frac{1}{r!}\nu^rP^r(f,g)
\end{equation}
where the first term is the ordinary pointwise product, $\displaystyle\nu=\frac{\hbar}{2\mathrm{i}}$ and in the succeeding terms, the bidifferential $P^r$ is expressed as
\begin{equation}\label{Bidifferential}
P^r(f,g)=\sum_{ij}\omega^{i_1j_1}\omega^{i_2j_2}\cdots\omega^{i_rj_r}\partial_{i_1i_2\cdots i_r}f\partial_{j_1j_2\cdots j_r}g.
\end{equation}
This $\star$-product provides a noncommutative but associative structure on $C^\infty(\Omega,\omega)$, and together with the Lie bracket
\begin{equation}
[f,g]_\nu=\frac{1}{2\nu}(f\star g-g\star f),
\end{equation}
with parameter $\nu^2$, the space $(C^\infty(\Omega)[[\nu]],\star,[\cdot,\cdot]_\nu)$ is a deformed algebra of classical observables. $(C^\infty(\Omega)[[\nu]]$ consists of formal power series in the parameter $\nu$ with coefficients in $C^\infty(\Omega).$ Just like any quantization procedure, we suppose that some Lie subalgebra $\widetilde{\mathfrak{g}}$ of $C^\infty(\Omega,\omega)$ is considered as a `preferred set of physical observables' preserved by deformation, that is, for any functions $a,b\in\widetilde{\mathfrak{g}},$ the following equation is satisfied
\begin{equation}\label{Covariance}
[a,b]_\nu=\{a,b\}.
\end{equation}	
A $\star$-product that satisfy equation (\ref{Covariance}) is said to be covariant and this property gives rise to a representation of a Lie group $G$ associated to $\widetilde{\mathfrak{g}}$ on $C^\infty(\Omega)[[\nu]]$ \cite{Arnal4}.	On the coadjoint orbit $\Omega_F$, the Lie algebra $\widetilde{\mathfrak{g}}$ are the energy functions $\widetilde{U}$ in (\ref{EnergyFunc}) and we can write equation (\ref{Covariance}) as 
\begin{equation}\label{Covariance1}
\frac{1}{2\nu}(\widetilde{U}\star\widetilde{T}-\widetilde{T}\star\widetilde{U})=\widetilde{[U,T]}.
\end{equation}

This covariance property is dependent on the choice of coordinate system on $\Omega$, since equation (\ref{Covariance}) suggests that the terms with parameter $\nu^2$ is eliminated only when the bidifferential expression $P^r(a,b)=0$ for $r>1$.

\subsection{Representations on the Lie algebra of observables}   

Suppose that the Moyal $\star$-product is a covariant $\star$-product. The left $\star$-product operators 
\begin{equation}\label{lOperator}
l_U(f)=\frac{1}{2\nu}\widetilde{U}\star f,f\in C^\infty(\Omega_F)[[\nu]]
\end{equation}	
define a Lie algebra representation of $\mathfrak{g}$ on $C^\infty(\Omega_F)[[\nu]]$; that is, $l$ is a linear map and preserves the Lie algebra structures between $\mathfrak{g}$ and the space of operators $l_U$. Indeed, the Moyal $\star$-product and the commutator bracket are bilinear,  so that
\begin{equation}
\widetilde{c_1U+c_2T}=c_1\widetilde{U}+c_2\widetilde{T}
\end{equation}
for any constants $c_1, c_2$ and $U,T\in\mathfrak{g}$, and using (\ref{Covariance1})
\begin{equation}
l_{[U,T]}(f)=\frac{1}{2\nu}\widetilde{[U,T]}\star f=[l_U,l_T](f),
\end{equation}	
for all $f\in C^\infty(\Omega_F)[[\nu]]$.

This representation of $\mathfrak{g}$ defined by (\ref{lOperator}) is the same type as the representation by derivation on the Schwartz class on the coadjoint orbit of a nilpotent Lie group $G$, as in \cite{Arnal2}. The Moyal $\star$-product of any two rapidly decreasing functions is rapidly decreasing and the operator (\ref{lOperator}) when defined on the Schwartz class on $\mathbb{R}^{2n}$ can be extended to a bounded linear operator on $L^2(\mathbb{R}^{2n})$ \cite{Arnal3}. These operators are intertwined with the partial Fourier transform
\begin{equation}
\mathcal{F}_p(f)(x,q)=\frac{1}{(2\pi)^n}\int_{\mathbb{R}^n}e^{-\mathrm{i}p\cdot x}f(p,q)\mathrm{d}p,
\end{equation}	 
that is, the operators
\begin{equation}
\hat{l}_U=\mathcal{F}_p\circ l_U\circ\mathcal{F}^{-1}_p
\end{equation}	
define a representation $\hat{l}$ of $\mathfrak{g}$ whose exponentiation $\exp \hat l$ gives the unitary representation of $G$.  

Although applied only to nilpotent and exponential Lie groups, these techniques has also worked well with concrete examples which are neither of the above-mentioned general classes.   
\section{Results}

For the 2-sphere as a result of the coadjoint action in (\ref{CoOrbit}) on the functional $F=(\mu,\alpha)\in\mathfrak{m}(3)^*$ when $\alpha=0$, we refer the reader to the work   \cite{Nable} regarding the unitary representations of $\mathrm{SO}(3)$ as a compact subgroup of $\mathrm{M}(3)$; via the projection $\mathrm{M}(3)\to\mathrm{SO}(3)$, these are the unitary representations of $\mathrm{M}(3)$ on the space of harmonic polynomials. In this work, we will illustrate the above-mentioned program on the 4-dimensional cotangent bundle of the 2-sphere when $\alpha\not= 0$.  

\subsection{Coordinate system on the cotangent bundle of a 2-sphere}
Although the spherical coordinate system will provide a global chart not only on the 2-sphere but as well as to its cotangent bundle, it is not used in the following computations. The energy function $\widetilde{U}$ on $\Omega_F$ defined by expression (\ref{EnergyFunc}) will contain the sine and cosine expressions under this coordinate system. But the sine and cosine is nonvanishing under differentiation. Hence, for any $U,T\in\mathfrak{m}(3)$,
\begin{equation}
\frac{1}{2\nu}(\widetilde{U}\star\widetilde{T}-\widetilde{T}\star\widetilde{U})=\widetilde{[U,T]}+o(\nu^2)
\end{equation}
since $P^r(\widetilde{U},\widetilde{T})=(-1)^rP^r(\widetilde{T},\widetilde{U})$. This coordinate system on $\Omega_F$ will not yield a covariant $\star$-product. Hence, to satisfy this covariant property of the Moyal $\star$-product, we choose a coordinate system on a flat neighborhood of an arbitrary point on $\Omega_F$. This local flatness is assured on a cotangent bundle since it is a symplectic manifold. 

Let $\Omega_F=T^*S^2_{\|\alpha\|}$ be the cotangent bundle on a 2-sphere with radius $\|\alpha\|$. Locally, we take $(p,q)$ or $p\mathrm{d}q$ as the usual coordinates of a neighborhood on $\Omega_F$ where $q\in S^2_{\|\alpha\|}$. With no loss of generality, we choose $q=(\|\alpha\|,0,0)$ and a locally flat neighborhood $O$ on $q$ since a 2-sphere is symplectic. The parameterization of $O$ will come from the tangent plane
\begin{equation*}
T_qS^2_{\|\alpha\|}=\mathrm{span}\{(0,\|\alpha\|,0),(0,0,\|\alpha\|)\}
\end{equation*}
via the exponential map. From elementary differential geometry, this exponential map $\exp:T_qS^2{\|\alpha\|}\to O$ is expressed via the integral curve $\gamma_v:I\to S^2_{\|\alpha\|}$, where $\gamma_v(0)=q$ and $\gamma_v'(0)=v\in T_qS^2_{\|\alpha\|},$ by $\exp(v)=\gamma_v(1)$. The integral curve is the geodesic
\begin{equation}\label{IntegralCurve}
\gamma_v(s)=\cos\left(\frac{\|v\|}{\|\alpha\|}s\right)q+\|\alpha\|\sin\left(\frac{\|v\|}{\|\alpha\|}s\right)\frac{v}{\|v\|}
\end{equation}	   
For computational ease, we set $\|v\|=1$, and sine and cosine expressions be equal to 1 in $\gamma_v(1)$. We identify $T_qS^2_{\|\alpha\|}$ with $\mathbb{R}^2$. From (\ref{IntegralCurve}), we define a local chart $\psi_1:\mathbb{R}^2\to O$ by
\begin{equation}\label{Parameter1}
\psi_1(t_1,t_2)=(\|\alpha\|,\|\alpha\|^2t_1,\|\alpha\|^2t_2).	
\end{equation}

There is a natural isomorphism between a tangent bundle and a cotangent bundle via the canonical symplectic structure of the cotangent bundle. Hence, the $p$ coordinate on the cotangent space at the position $q$ can be identified with the vector coordinate on the tangent space at the same position $q$. Let $\varphi=p\mathrm{d}q$. From (\ref{Parameter1}), we write $\varphi$ as
\begin{equation}
\varphi=\|\alpha\|^2p_2\mathrm{d}t_1+\|\alpha\|^2p_3\mathrm{d}t_2.
\end{equation}	 
Let $s_1=\|\alpha\|^2p_2$ and $s_2=\|\alpha\|^2p_3$. Hence, we define a local chart $\psi_2:\mathbb{R}^4\to T^*O$ by
\begin{equation}\label{Parameter2}
\psi_2(s_1,s_2,t_1,t_2)=\left(0,\frac{s_1}{\|\alpha\|^2},\frac{s_2}{\|\alpha\|^2},\|\alpha\|,\|\alpha\|^2t_1,\|\alpha\|^2t_2\right).
\end{equation}	 
Given this chart in (\ref{Parameter2}), any point $F'$ in $T^*O$ is expressed as
\begin{equation}
F'=\frac{s_1}{\|\alpha\|^2}X_2^*+\frac{s_2}{\|\alpha\|^2}X_3^*+\|\alpha\|E_1^*+\|\alpha\|^2t_1E_2^*+\|\alpha\|^2t_2E_3^*.
\end{equation}	 
At this point, it is understood that all functions on $\Omega_F$ are parameterized by the coordinates $(s,t)=(s_1,s_2,t_1,t_2)$. We write $f(s,t)$ instead of $(f\circ\psi_2)(s,t)$.

\subsection{Hamiltonian system on $T^*O$ and covariant $\star$-product}

On $T^*O\subset T^*S^2_{\|\alpha\|}$, the energy function $\widetilde{U}:T^*O\to\mathbb{R}$ (\ref{EnergyFunc}) is given by
\begin{equation}\label{EnergyFunc1}
\widetilde{U}(s,t)=\frac{x_2}{\|\alpha\|^2}s_1+\frac{x_3}{\|\alpha\|^2}s_2+\|\alpha\|e_1+\|\alpha\|^2e_2t_1+\|\alpha\|^2e_3t_2
\end{equation}
for each $U=x_1X_1+x_2X_2+x_3X_3+e_1E_1+e_2E_2+e_3E_3\in\mathfrak{m}(3)$. The Hamiltonian vector field associated to $\widetilde{U}$ is
\begin{equation}
\xi_U=\left(\frac{x_2}{\|\alpha\|^2}\frac{\partial}{\partial t_1}-\|\alpha\|^2e_2\frac{\partial}{\partial s_1}\right)+\left(\frac{x_3}{\|\alpha\|^2}\frac{\partial}{\partial t_2}-\|\alpha\|^2e_3\frac{\partial}{\partial s_2}\right).
\end{equation}
Using equation (\ref{Symplectic}), the Kirillov symplectic form  at $F=\|\alpha\|E^*_1$ is
\begin{equation}
\omega_F=\|\alpha\|(\mathrm{d}t_2\wedge\mathrm{d}s_1+\mathrm{d}s_2\wedge\mathrm{d}t_1).
\end{equation}
The $\omega^{ij}$ expression in the  bidiffirential expression $P^r$ in (\ref{Bidifferential}) are the entries of the matrix
\begin{equation}
\omega_F=\left(\begin{array}{cccc}
0 & 0 & 0 & -1\\
0 & 0 & 1 & 0 \\
0 & -1 & 0 & 0 \\
1 & 0 & 0 & 0
\end{array}	\right)
\end{equation}

With this parameterization on $T^*O$, the bidifferential $P^r(\widetilde{U},\widetilde{T})=0$ for $r>1$ since $\widetilde{U}$ is linear. So,
\begin{equation}\label{CovariantExample}
\widetilde{U}\star\widetilde{T}-\widetilde{T}\star\widetilde{U}=\frac{\hbar}{\mathrm{i}}P^1(\widetilde{U},\widetilde{T}).
\end{equation}
Since
\begin{equation}
P^1(\widetilde{U},\widetilde{T})=\omega_F(\xi_U,\xi_T),
\end{equation}
together with (\ref{Symplectic}) at $F=\|\alpha\|E^*_1$, equation (\ref{CovariantExample}) is exactly (\ref{Covariance1}). Hence, the Moyal $\star$-product is covariant. In the next section, the representation $l$ which is defined by the operators (\ref{lOperator}) will be used in the computation of the unitary representation of $\mathrm{M}(3)$. For simplicity, we set $\hbar=1$.  

\subsection{Unitary representations of $\mathrm{M}(3)$}

The aim of this section is to compute for the concrete expression of the operator $\hat{l}_U=\mathcal{F}_p\circ l_U\circ\mathcal{F}^{-1}_p$ and its exponentiation, which is the desired unitary representation of $\mathrm M(3)$. The partial Fourier transform on the momentum variable $p=(s_1,s_2)$ in these computations is defined by the expression
\begin{equation}
(\mathcal{F}_sf)(\eta,t)=\frac{1}{2\pi}\int_{\mathbb{R}^2}\mathrm{e}^{-\mathrm{i}(s_1\eta_1+s_2\eta_2)}f(s_1,s_2,t_1,t_2)\mathrm{d}s_1\mathrm{d}s_2
\end{equation}	 
and its inverse as 
\begin{equation}
(\mathcal{F}^{-1}_sf)(s,t)=\frac{1}{2\pi}\int_{\mathbb{R}^2}\mathrm{e}^{\mathrm{i}(s_1\eta_1+s_2\eta_2)}f(\eta_1,\eta_2,t_1,t_2)\mathrm{d}\eta_1\mathrm{d}\eta_2.
\end{equation}
Let $f$ be a rapidly decreasing function on $\Omega_F$. For
\begin{equation}\label{37}
\hat{l}_U(f)=\mathrm{i}\mathcal{F}_s\left(\widetilde{U}\star\mathcal{F}^{-1}_sf\right),
\end{equation}
we have
\begin{equation}\label{38}
\widetilde{U}\star\mathcal{F}^{-1}_s(f)=\widetilde{U} \mathcal{F}^{-1}_sf+\frac{1}{2\mathrm{i}}P^1\left(\widetilde{U},\mathcal{F}^{-1}_sf\right),
\end{equation}
where the first term is the pointwise product of the energy function in (\ref{EnergyFunc1}) and $\mathcal{F}^{-1}_sf$ while the Poisson bracket in the second term is
\begin{eqnarray}
P^1\left(\widetilde{U},\mathcal{F}^{-1}_sf\right)&=&\mathrm{i}\|\alpha\|^3\left[e_3\mathcal{F}^{-1}_s(\eta_1f)-e_2\mathcal{F}^{-1}_s(\eta_2f)\right]\\ \nonumber &&+\frac{1}{\|\alpha\|}\left[x_3\mathcal{F}^{-1}_s\left(\frac{\partial f}{\partial t_1}\right)-x_2\mathcal{F}^{-1}_s\left(\frac{\partial f}{\partial t_2}\right)\right].
\end{eqnarray}
Furthermore, when we apply the partial Fourier transform in each of the terms in (\ref{38}), we have
\begin{eqnarray}\label{40}
\mathcal{F}_s\left(\widetilde{U}\mathcal{F}^{-1}_sf\right)&=&\frac{x_2}{\|\alpha\|^2}\mathcal{F}_s\left(s_1\mathcal{F}^{-1}_sf\right)+\frac{x_3}{\|\alpha\|^2}\mathcal{F}_s\left(s_2\mathcal{F}^{-1}_sf\right)\\ \nonumber && +\left(\|\alpha\|e_1+\|\alpha\|^2e_2t_1+\|\alpha\|^2e_3t_2\right)f
\end{eqnarray}	 
and
\begin{eqnarray}\label{41}
\mathcal{F}_s\left(P^1\left(\widetilde{U},\mathcal{F}^{-1}_sf\right)\right)&=&\mathrm{i}\|\alpha\|^3\left[e_3\eta_1-e_2\eta_2\right]f\\ \nonumber &&+\frac{1}{\|\alpha\|}\left[x_3\frac{\partial}{\partial t_1}-x_2\frac{\partial}{\partial t_2}\right]f.
\end{eqnarray}	
It is easy to show that $\mathcal{F}_s(s_jf)=\mathrm{i}\partial_{\eta_j}\mathcal{F}_sf$. Applying this in (\ref{40}) and combining the outcome with (\ref{41}), the operator (\ref{37}) becomes
\begin{eqnarray}
\hat{l}_U(f)&=&\mathrm{i}\mathcal{F}_s\left(\widetilde{U}\mathcal{F}^{-1}_sf\right)+\frac{1}{2}\mathcal{F}_s\left(P^1\left(\widetilde{U},\mathcal{F}^{-1}_sf\right)\right)\\
&=&\mathrm{i}\|\alpha\|\left[e_1+e_2\left(\|\alpha\|t_1-\frac{\|\alpha\|^2}{2}\eta_2\right)+e_3\left(\|\alpha\|t_2+\frac{\|\alpha\|^2}{2}\eta_1\right)\right]f \nonumber\\
&&+\left[x_3\left(\frac{1}{2\|\alpha\|}\frac{\partial}{\partial t_1}-\frac{1}{\|\alpha\|^2}\frac{\partial}{\partial\eta_2}\right)-x_2\left(\frac{1}{2\|\alpha\|}\frac{\partial}{\partial t_2}+\frac{1}{\|\alpha\|^2}\frac{\partial}{\partial\eta_1}\right)\right]f.\nonumber
\end{eqnarray}
The expression above can be simplified by some change of variable. If we let $u=\|\alpha\|t_1-\frac{\|\alpha\|^2}{2}\eta_2$ and $v=\|\alpha\|t_2+\frac{\|\alpha\|^2}{2}\eta_1$, then the operator $\hat{l}_U$ becomes
\begin{equation}\label{lU1}
\hat{l}_U=\mathrm{i}\|\alpha\|(e_1+e_2u+e_3v)-\left(x_2\frac{\partial}{\partial v}-x_3\frac{\partial}{\partial u}\right).
\end{equation}

The local operator (\ref{lU1}) defines a representation of the Lie algebra $\mathfrak{m}(3)$ on $L^2(T^*O)$ where we have defined $O$ as a flat neighborhood, centered at $q=(\|\alpha\|,0,0)$. When $O$ is centered at any $q$ in $S^2_{\|\alpha\|}$, the above-computed operator can be categorically expressed as   operators of the form
\begin{equation}\label{lU2}
\hat{l}_{E_i}=\mathrm{i}\|\alpha\|s_i\quad\text{ and }\quad\hat{l}_{X_j}=-\frac{\partial}{\partial s_j},
\end{equation}	
for $i,j=1,2,3$ where $s\in S^2_{\|\alpha\|}$. The simply-connectedness of the 2-sphere assures us that these local operators are global operators via the Monodromy Theorem. Hence, the operator $\hat{l}_U$ acts on functions on the whole coadjoint orbit. 

As observed, we can see the similarity of the position and momentum operators in (\ref{Position}) and (\ref{Momentum}), respectively, where $\lambda=\|\alpha\|$. To prove that the $l_{E_i}$ and $l_{X_j}$ are the operators that defines the infinitisimal representation of $\mathfrak{m}(3)$, let us first replace the group elements of $\mathrm{M}(3)$ with 1-parameter subgroups $\exp e_iE_i$ and $\exp x_jX_j$ on the unitary operators in (\ref{UnitRep}) and these are
\begin{equation}\label{45}
(\mathcal{U}^\lambda_{\exp e_iE_i}f)(s)=\mathrm{e}^{\mathrm{i}\lambda e_is_i}f(s)
\end{equation}	    
and
\begin{equation}\label{46}
(\mathcal{U}^\lambda_{\exp x_jX_j}f)(s)=f((\exp -x_jX_j)s).
\end{equation}
In equation (\ref{46}), the action of the exponential $\exp -x_jX_j$ on $s$ rotates the plane perpendicular to the $j$th axis; hence, an angular translation by some $-x_j$ units. Using the Taylor series expansion of this translation, we have 
\begin{equation}\label{47}
(\mathcal{U}^\lambda_{\exp x_jX_j}f)(s)=\mathrm{e}^{-x_j\frac{\partial}{\partial s_j}} f(s).
\end{equation}
The derivatives of (\ref{45}) and (\ref{47}) with respect to their parameters are exactly the actions of the operators $\hat{l}_{E_i}$ and $\hat{l}_{X_j}$, that is,
\begin{equation}\label{48}
\frac{\mathrm{d}}{\mathrm{d}e_i}\left(\mathcal{U}^\lambda_{\exp e_iE_i}f\right)(s)=(\mathrm{i}\lambda s_i)\mathrm{e}^{\mathrm{i}\lambda e_is_i}f(s)=\hat{l}_{E_i}\left(\mathcal{U}^\lambda_{\exp e_iE_i}f\right)(s)
\end{equation}	
and
\begin{equation}\label{49}
\frac{\mathrm{d}}{\mathrm{d}x_j}\left(\mathcal{U}^\lambda_{\exp x_jX_j}f\right)(s)=-\mathrm{e}^{-x_i\frac{\partial}{\partial s_j}}\frac{\partial}{\partial s_j}f(s)=\hat{l}_{X_j}\left(\mathcal{U}^\lambda_{\exp x_jX_j}f\right)(s).
\end{equation}
Since the initial values of (\ref{48}) and (\ref{49}) are respectively $\left.\left(\mathcal{U}^\lambda_{\exp e_iE_i}f\right)(s)\right|_{e_i=0}=f(s)$ and $\left.\left(\mathcal{U}^\lambda_{\exp x_jX_j}f\right)(s)\right|_{x_j=0}=f(s)$, the expressions $\left(\mathcal{U}^\lambda_{\exp e_iE_i}f\right)(s)$ and $\left(\mathcal{U}^\lambda_{\exp x_jX_j}f\right)(s)$ are the respective unique solutions to the Cauchy problem
\begin{equation}
\left\{\begin{array}{r c l}
\displaystyle\frac{\mathrm{d}}{\mathrm{d}t}T(t,s)&=&\hat{l}_UT(t,s)\vspace{3mm}\\ 
T(0,s)&=&\mathrm{Id}
\end{array}	  \right. .
\end{equation}	
Hence, the exponentiation of the operators $\hat{l}_{E_i}$ and $\hat{l}_{X_j}$ are the unitary operators that generate the unitary representation of $\mathrm{M}(3)$. Via the Baker-Campbell-Hausdorff formula, we recover the operators of the unitary representation $\mathcal{U}^\lambda$ of $\mathrm{M}(3)$ by
  \begin{eqnarray}
\mathcal{U}^\lambda_g=\exp(e_1\hat{l}_{E_1}+e_2\hat{l}_{E_2}+\hat{l}_{E_3})\times\exp(x_1\hat{l}_{X_1})\exp(x_2\hat{l}_{X_2})\exp(x_3\hat{l}_{X_3}),
\end{eqnarray}	 	
classified by the radius $\lambda=\|\alpha\|$ of the coadjoint orbit $\Omega_F$.

\subsection{Star-polarization}

On $L^2(\Omega_F)$, the unitary representation $\mathcal{U}^\lambda$ is not irreducible. The method of $\star$-polarization reduces this space to a more physically acceptable  space of functions; as defined, these functions, say $f$, must satisfy the equation
\begin{equation}\label{polarization}
f\star a=\chi(a)f
\end{equation}	 
where $\chi$ is a character on a subalgebra $\widetilde{\mathfrak{g}}_0$ of $\widetilde{\mathfrak{g}}$ \cite{Fronsdal}. This collection of functions $f$ is in fact a topological subspace of $C^\infty(\Omega_F)$ which is stable and irreducible under $f\mapsto \frac{1}{2\nu}\widetilde{U}\star f$, for all $U\in\mathfrak{g}$. This method is a generalization of the polarization of Kostant \cite{Kostant}.

In this computation, we let $\widetilde{\mathfrak{g}}_0$ be the subalgebra spanned by the parameters associated the commutative vectors $E_i$'s of $\mathfrak{m}(3)$. Given our paramterization $(s,t)$ on $T^*O$, let $\widetilde{E}_i=\|\alpha\|^2e_it_i$ from (\ref{EnergyFunc1}) and $\chi(\widetilde{E}_i)=\|\alpha\|^2e_i\chi_i$ where $\chi_i\in\mathbb{C}^*$. Equation $(\ref{polarization})$ is expressed as
\begin{equation}
f\star\widetilde{E}_i=\chi(\widetilde{E}_i)f.
\end{equation}
A straightforward calculation simplifies this equation into an ordinary differential equation
\begin{equation}\label{ODE}
\frac{\|\alpha\|}{2\mathrm{i}}\frac{\partial f}{\partial s_j}=(t_i-\chi_i)f
\end{equation}
for all $i=1,2,3$. The solutions to (\ref{ODE}) are functions having the form
\begin{equation}
f_\chi(s,t)=\mathrm{e}^\frac{\mathrm{i}2\left[s_2(t_1-\chi_1)+s_1(t_2-\chi_2)\right]}{\|\alpha\|}\psi(t)
\end{equation}
where $\psi$ is defined on $O$. But $f_\chi$ is in $L^2(O)$, only if $f$ is in $L^2(T^*O)$ \cite{Huynh} where
\begin{equation}
f=\int f_\chi\mathrm{d}\chi.
\end{equation}
Hence via the Monodromy Theorem, $\exp\hat{l}_U$ are operators on $L^2(S^2_{\|\alpha\|})$, or simply on $L^2(S^2)$, for all $U\in\mathfrak{m}(3)$.

\section{Conclusion}

This paper has shown that the unitary representation of the Euclidean motion group $\mathrm{M}(3)$ can be constructed from the Moyal $\star$-product quantization. We mention also that, as a byproduct, a quantum algebra of functions on the tangent bundle of spheres has been constructed. Unitary representation theory has an important role in the mathematical formalism of quantum mechanics, but these computations exhibit the converse. That is, deformation quantization where the Moyal $\star$-product as an example has contributed to the solution to the basic question of unitary representation theory. This correspondence between quantum theory and unitary representations of Lie groups has been present since the early days of quantum mechanics, and has enriched both the physics and mathematics literature.

Though this paper does not claim that the presented procedure will be always true for all Lie groups, however, it may be true for coadjoint orbits of semi-direct products of a compact Lie group with a vector space. This paper provides this hint but needs to be formalized. 

\section*{Acknowledgment and Dedication}
The first author expresses its gratitude to the Faculty Development Program of the Philippines' Commission on Higher Education for the financial support it has given and to the Ateneo de Manila University where part of this work was conducted as a graduate student.

This second author dedicates this work to Prof. Do Ngoc Diep of the Institute of Mathematics, Vietnam Academy of Science and Technology.

\noindent Alexander J. Balsomo \\
Department of Mathematics \\
West Visayas State University\\
Iloilo City, Philippines\\
{\it E-mail address}: {\tt abalsomo@wvsu.edu.ph}\\[0.3cm]

\noindent Job A. Nable \\
Department of Mathematics \\
Ateneo de Manila University\\
Quezon City, Philippines\\
{\it E-mail address}: {\tt jnable@ateneo.edu}

\end{document}